\documentclass[reprint,amsmath,amssymb,aps,pra,superscriptaddress,nofootinbib]{revtex4-2}

\usepackage[utf8]{inputenc}
\usepackage[page]{appendix}
\usepackage[english]{babel}
\usepackage{hyperref}
\usepackage{hyperref}
\usepackage{physics}
\usepackage{float}
\usepackage{graphicx}
\usepackage[capitalise]{cleveref}
\usepackage{mhchem}
\usepackage{enumitem}
\usepackage[dvipsnames]{xcolor}
\usepackage[
    protrusion=true,
    activate={true,nocompatibiality},
    final,
    tracking=true,
    kerning=true,
    spacing=true,
    factor=1000]{microtype}
\hypersetup{
 colorlinks=true, 
 linkcolor=blue, 
 citecolor=blue,        
 filecolor=blue,      
 urlcolor=blue}

\usepackage{titlesec}
\titleformat*{\section}{\bfseries\large}
\titleformat*{\subsection}{\bfseries}
\usepackage{caption}

\newcommand{\mr}[1]{\mathrm{#1}}

\Crefformat{appendix}{Supplementary Information #2#1#3}
\begin{document}
                                                         
\author{Gr\'egory Moille}
\email{gregory.moille@nist.gov}
\affiliation{Joint Quantum Institute, NIST/University of Maryland, College Park, USA}
\affiliation{Microsystems and Nanotechnology Division, National Institute of Standards and Technology, Gaithersburg, USA}
\author{Pradyoth Shandilya}
\affiliation{University of Maryland, Baltimore County, Baltimore, MD, USA}
\author{Alioune Niang}
\affiliation{University of Maryland, Baltimore County, Baltimore, MD, USA}
\author{Curtis Menyuk}
\affiliation{University of Maryland, Baltimore County, Baltimore, MD, USA}
\author{Gary Carter}
\affiliation{University of Maryland, Baltimore County, Baltimore, MD, USA}
\author{Kartik Srinivasan}
\email{kartik.srinivasan@nist.gov}
\affiliation{Joint Quantum Institute, NIST/University of Maryland, College Park, USA}
\affiliation{Microsystems and Nanotechnology Division, National Institute of Standards and Technology, Gaithersburg, USA}
\date{\today}

\newcommand{\mytitle}{Versatile Optical Frequency Division with Kerr-induced Synchronization at Tunable Microcomb Synthetic Dispersive Waves}
\title{\mytitle}

\begin{abstract} 
    Kerr-induced synchronization (KIS) provides a key tool for the control and stabilization of a dissipative Kerr soliton (DKS) frequency comb, enabled by the capture of a comb tooth by an injected reference laser.
    Efficient KIS relies on large locking bandwidth, meaning both the comb tooth and intracavity reference power need to be sufficiently large. %
    Although KIS can theoretically occur at any comb tooth, large modal separations from the main pump to achieve large optical frequency divisions are often difficult or unfeasible due to cavity dispersion. While tailoring the dispersion to generate dispersive waves (DWs) can support on-resonance KIS far from the main pump, this approach restricts synchronization to specific wavelengths. 
    Here, we demonstrate an alternative KIS method that allows efficient synchronization at arbitrary modes by multi-pumping a microresonator. This creates a multi-color DKS with a main and an auxiliary comb, the latter enabling the creation of a synthetic DW. As cross-phase modulation leads to a unique group velocity for both the soliton comb and the auxiliary comb, repetition rate disciplining of the auxiliary comb through KIS automatically controls the DKS microcomb. We explore this color-KIS phenomenon theoretically and experimentally, showing control and tuning of the soliton microcomb repetition rate, resulting in optical frequency division independent of the main pump noise properties.
\end{abstract}
\maketitle

\begin{figure*}
    \begin{center}    
    \includegraphics{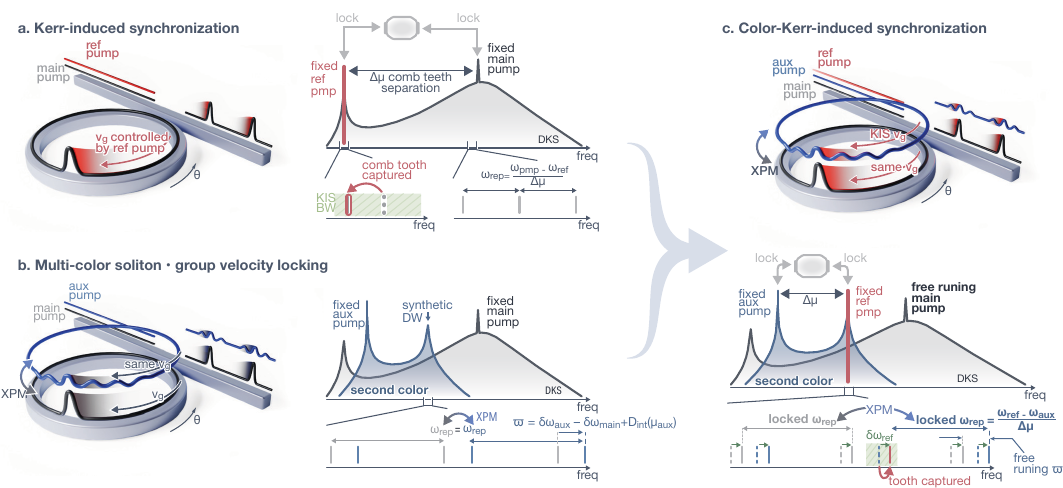}
    \end{center}
    \caption{\label{fig:1} \textbf{Concept of color-Kerr-induced synchronization. }
    \textbf{a} Kerr-induced synchronization (KIS) concept for optical frequency division (OFD). A dissipative Kerr soliton is generated by driving a microring resonator with the main pump propagating in the clockwise direction of the azimuthal dimension of the microresonator (\textit{i.e.} along $-\theta$). An injected reference laser captures the nearest comb tooth. Any variation of frequency of the two locked pumps will be divided onto the repetition rate of the DKS by a factor $\Delta\mu$ being the comb tooth separation between the two pumps, and hence the DKS group velocity is controlled through the reference pump. %
    \textbf{b} Outside of synchronization, the reference pump does not capture a comb tooth. However, cross-phase modulation (XPM) binds the intracavity color triggered by this auxiliary pump, making it travel at the same group velocity as the DKS. It results in a two-component comb sharing the exact same repetition rate, yet offset in carrier envelope offset from one another by $\varpi = \delta \omega_\mathrm{aux}-\delta\omega_\mathrm{main} + D_\mathrm{int}(\mu_\mathrm{aux})$, accounting for the main and auxiliary pump detunings and the integrated dispersion at the auxiliary pumped mode. In this case, the repetition rate is not disciplined through OFD. However, a new phase-matching condition occurs, where a synthetic dispersive wave is created.
    \textbf{c} In color-KIS, we harness both effects. The secondary color's creation of a DW enables both large comb tooth power and an on-resonance reference pump for efficient KIS -- this time at the secondary color. However, as XPM binds the two colors together, any tuning of the repetition rate through the frequency tuning of the reference $\delta\omega_\mathrm{ref}$ that is imprinted onto the secondary color is automatically replicated onto the DKS. Hence, the auxiliary and reference pump can be independently optically locked for OFD, while the main DKS pump remains free running, since the relative comb color offset $\varpi$ absorbs any frequency noise from the main pump.
    }
\end{figure*}

\section{Introduction}
Optical frequency combs (OFCs) are, in their low noise state, an essential component in the metrology toolbox for frequency and time measurement~\cite{FortierCommunPhys2019a,DiddamsScience2020a}. They can be used as a phase-coherent frequency multiplier/divider between the microwave and optical domains for optical frequency synthesis~\cite{SpencerNature2018}, optical clockworks~\cite{NewmanOptica2019, OelkerNat.Photonics2019}, time transfer~\cite{GiorgettaNaturePhoton2013,CaldwellNature2023}, and ultra-low noise microwave generation~\cite{TetsumotoNat.Photon.2021, XieNaturePhoton2017}, while also increasingly finding application in more widespread consumer technologies such as distance ranging~\cite{RiemensbergerNature2020}. Integrating such OFCs on chip has been a recurrent quest over the last decade, in particular with the demonstration of dissipative Kerr soliton (DKS) microcombs which convert a continuous wave (CW) pump into a periodically-extracted pulse train~\cite{KippenbergScience2018}. DKSs at their core rely on the dispersion compensation by Kerr nonlinearity present in several CMOS-compatible materials, such as silicon nitride (\ce{Si3N4}), allowing for low-cost and mass-scale fabrication of high quality factor microring~\cite{LiuNatCommun2021a} or Fabry-Perot resonators~\cite{WildiOpticaOPTICA2023} through which DKSs can be generated with low pump power~\cite{SternNature2018b}.
Although the combs' technical requirements may vary from one application to another, the control and/or stabilization of their repetition rate is in general a key element~\cite{DiddamsScience2020a}. In particular, optical frequency division (OFD) relies on pinning two comb teeth to stable lasers, leveraging the fundamental property that their noise will be divided onto the repetition rate by their comb tooth separation~\cite{FortierNaturePhoton2011a}. Although on-chip operation for the laser stabilization~\cite{LiuOpticaOPTICA2022} and the comb generation~\cite{Kudelin2023, Sun2023} can be achieved, the overall architecture mostly relies on lab-scale equipment for the comb-laser locking with only one degree of freedom (i.e. the main pump). Recently, Kerr-induced synchronization (KIS) has been demonstrated as an all-optical approach to address the technical challenge of optically locking the comb~\cite{MoilleNature2023}. In KIS, the injection of another reference pump laser into the comb-generating resonator results in the closest comb tooth snapping onto the reference [\cref{fig:1}(a)], enabling external control of two comb teeth through both pumps. This method  has been successfully shown to enable all-optical OFD through locking of both pumps~\cite{MoilleNature2023, WildiAPLPhotonics2023}, with performance on par with other techniques~\cite{Kudelin2023, Sun2024}. 

Although KIS greatly simplifies an optical clockwork~\cite{MoilleNature2023, MoilleFront.Opt.2023}, it can be challenging to implement for other OFD applications. First, it fixes the main pump, rendering the system stiff if low phase-noise is to be achieved, and greatly limiting the choice of frequency for the reference laser that provides the second pinning point for the comb. Second, self-injection locking (SIL), which has enabled turn-key DKS generation~\cite{ShenNature2020b, UlanovNat.Photon.2024, VoloshinNatCommun2021a}, will in most cases only be possible at the main pump, and not at the reference pump, since the cavity dispersion creates a frequency shift between the comb line and its nearest resonance~\cite{MoilleNat.Commun.2021a}. This frequency offset, when larger than the KIS bandwidth, means that resonant feedback for SIL will not available at a frequency for which KIS can occur, thereby limiting the ability to realize a system that exploits dual-SIL for turn-key DKS generation and OFD.
 Third, this KIS phase locking efficiency relies on the KIS coupling energy, which can be expressed as the geomerical mean of the intra-cavity power of the comb tooth and the reference pump upon synchronization~\cite{MoilleNature2023,WildiAPLPhotonics2023}. The injection of the reference laser at a dispersive wave (DW) enhances the KIS energy since the comb tooth is resonant, maximizing both the tooth and reference intracavity power. However, DWs are not always present, especially in microcombs with directly detectable, low repetition rates that are often achieved in resonators with pure quadratic dispersion. It has been shown that DW operation is not a prerequisite for KIS~\cite{WildiAPLPhotonics2023}. However it is generally advantageous for the dual-pinning to occur between widely separated comb teeth, to provide a large OFD factor. Here, the $sech^2$ microcomb envelope greatly reduces the available comb tooth power far from the main pump, while the larger spectral offset between comb tooth and resonance (in absence of a DW) leads to poor injection of the reference into the resonator. These effects may lead to a KIS bandwidth narrower than the laser linewidth, making synchronization practically unachievable. Here we propose a new scheme that uncouples the main pump soliton -- which can remain free-running -- with the all-optical OFD enabled through KIS, by using a multi-color cavity soliton in a microring resonator~\cite{MoilleNat.Commun.2021a}. 

\begin{figure*}
    \begin{center}
        \includegraphics[width = \textwidth]{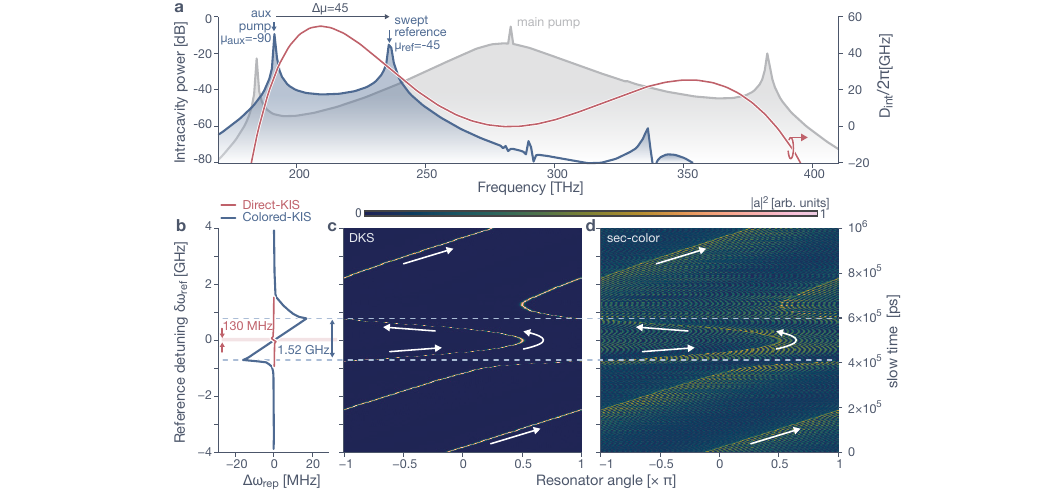}
        \caption{\label{fig:2}%
        \textbf{Theoretical study of the color-KIS effect.} %
        \textbf{a} Two-wave LLE simulation results for the DKS (gray) and secondary color (blue), using the dispersion profile similar to experimental system (red, right scale). The auxiliary pump is held fixed and on-resonance at $\mu=-90$. The reference pump, which is injected at the (high frequency) second-color DW, is adiabatically swept. The power is normalized to 1~mW. %
        \textbf{b} Repetition rate variation of the DKS $\Delta\omega_\mathrm{rep}$ against time, with a linear ramp of the reference pump detuning $\delta\omega_\mathrm{ref}$. Once close enough to the secondary-color comb tooth, the reference pump grabs it and modifies the secondary color repetition rate $\omega_\mathrm{rep}$ (blue). Through XPM, the DKS repetition rate is also modified. We compare it against direct KIS at the same mode $\mu_\mathrm{ref} = -45$ (red), resulting in an opposite $\omega_\mathrm{rep}$ slope, with a KIS bandwidth (BW) about 10$\times$ smaller than the color-KIS BW with the same reference pump power. The detuning $\delta\omega_\mathrm{ref}$ in both cases is normalized such that a null repetition rate variation while synchronized corresponds to the same repetition rate when outside the synchronization regime. 
        \textbf{c} and \textbf{d} show evolution of the DKS and secondary color in the resonator angle against the slow time $t$ (where $\approx$~1~ps corresponds to one soliton round trip), hence the reference detuning. Their trajectories follow the same path as their group velocities are bound, even when the synchronization occurs at the secondary color. Outside of the color-KIS, their drift in the slow time $t$ is the same highlighting the absence of impact of the reference onto the soliton group velocity. In the second-color synchronization regime, the drift becomes dependent with the reference detuning, hence, color-KIS provides a means to change the DKS repetition rate. %
        }
    \end{center}
    \vspace{-1em}
\end{figure*}

\begin{figure*}
    \begin{center}
        \includegraphics[width = \textwidth]{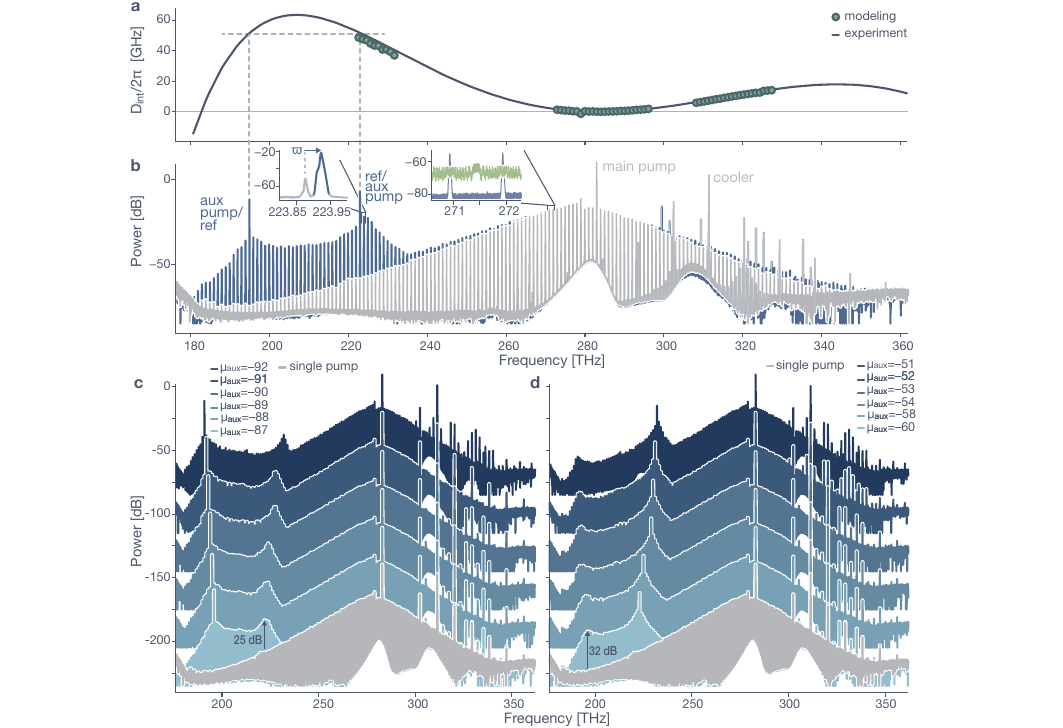}
        \caption{\label{fig:3}
        \textbf{DKS and new DW generated via auxiliary pumping.} %
        \textbf{a} Integrated dispersion from wavemeter-calibrated resonance frequencies (50~MHz uncertainty, within marker size) measurement of the fundamental transverse electric mode (circles) and from the finite-element method dispersion modeling (solid line), showcasing close agreement and used in \cref{fig:2}. The phase matching with the auxiliary pump for where creation of a new synthetic DW appears is shown in the gray dashed line.
        \textbf{b} Single-pumped DKS microcomb spectrum (gray) obtained using the counterpropagative cooler method~\cite{ZhangNatCommun2020, ZhouLightSciAppl2019}. The secondary-color (blue) is generated by the auxiliary pump, creating a new synthetic DW where the reference pump can be sent to trigger the color-KIS. The DKS and secondary color are offset from each other by about $\varpi = +40$~GHz (left inset, the grey comb line is present in either the single pump or auxiliary pump case, while the blue comb line originates from the auxiliary pump), in accordance with the integrated dispersion value at the auxiliary pump. For subsequent measurements in Fig.~\ref{fig:4}, we measure the repetition rate of only the DKS using an electro-optic comb apparatus that modulates two adjacent DKS comb teeth (right inset; green is the electro-optically modulated spectrum). The power is normalized to 1~mW. %
        \textbf{c} and \textbf{d} are optical spectra for different auxiliary pumped modes in the 190~THz band and 230~THz band, respectively. Compared to the single-pumped DKS (gray), the available comb tooth power for KIS is increased by 25~dB and 32~dB, respectively.
        }
    \end{center}
\end{figure*}

\begin{figure*}
    \begin{center}
        \includegraphics[width = \textwidth]{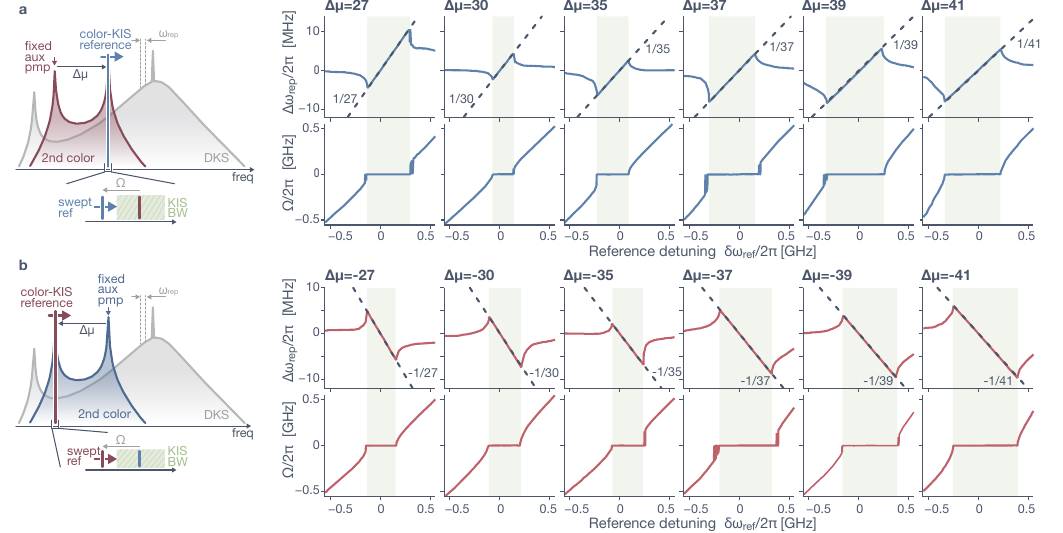}
        \caption{\label{fig:4}
        \textbf{Repetition rate control and optical frequency division through color-KIS.} Characterization of the repetition rate entrainment (top) and beat note between the reference and secondary color (bottom) for different mode spacing $\Delta \mu$ between auxiliary and reference pumps, corresponding to the different auxiliary pumping presented in~\cref{fig:3}. The repetition rate $\omega_\mathrm{rep}$ is only measured at the DKS. Both $\Delta \omega_\mathrm{rep}$ and $\Omega$ are extracted from simultaneous temporal measurement and calibrated according to Supplementary Information S.2. The color-KIS regime is highlighted in the green area, where a linear change of $\omega_\mathrm{rep}$ following the $\Delta \mu$ OFD factor is observed (dashed lines). We emphasize that this trend is neither offset nor scaled thanks to accurate measurement of the zero detuning. Simultaneously, $\Omega$ is locked to zero and hence the reference pump is capturing its nearest secondary-color comb tooth. 
        \textbf{a} Measurements in the case of a fixed auxiliary pump in the 190~THz band and a swept reference pump in the 230~THz range. 
        \textbf{b} Measurements in the case of a fixed auxiliary pump in the 230~THz band and a swept reference pump in the 190~THz range. The change in sign of the slope of $\Delta\omega_\mathrm{rep}$ within the color-KIS window between \textbf{a} and \textbf{b} is consistent with the change in sign of $\Delta\mu$.
        }
    \end{center}
\end{figure*}

\begin{figure}
    \begin{center}
        \includegraphics{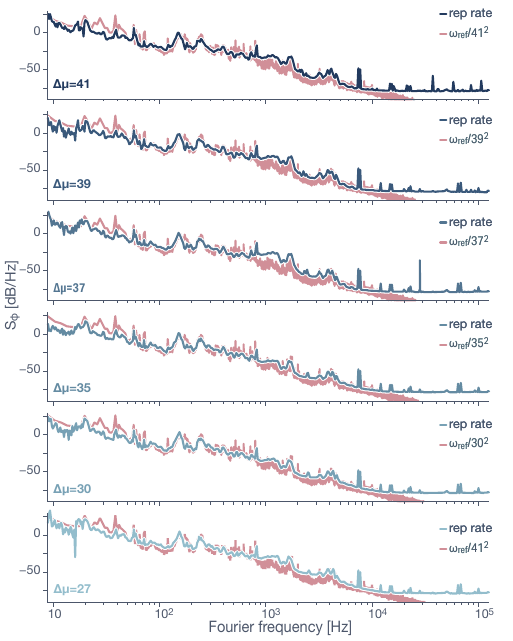}
        \caption{\label{fig:5} \textbf{Color-KIS noise optical frequency division onto the DKS repetition rate}.
        Optical frequency division characterization of the repetition rate phase noise for different values of $\Delta\mu$. The auxiliary laser and reference laser are both locked to separate 10~MHz free-spectral range Mach-Zehnder interferometers, with the 230~THz range CTL carrying most of the optical noise. Once divided by the appropriate OFD factor $\Delta \mu^2$, the laser phase noise (shown in red) overlaps with the DKS $\omega_\mathrm{rep}$ phase noise (shown in dark blue to light blue for decreasing $\Delta\mu$) according to the KIS principle. Here, the stability of the DKS repetition rate comes from KIS of $\omega_\mathrm{rep}$ of the secondary-color, which is then transferred to the DKS $\omega_\mathrm{rep}$ through XPM. The power is referenced to that of the carrier, namely, dBc/Hz
        }
    \end{center}
\end{figure}

\section{Main Text}
\subsection{Physical concept of the color-KIS} 
We start by reviewing multi-pumped DKS operation outside of synchronization. In this case, the second injected laser does not become a DKS comb tooth [\cref{fig:1}(b)]. Instead, it stimulates the formation of an auxiliary comb~\cite{MoilleNat.Commun.2021a,ZhangNatCommun2020}, with cross-phase modulation (XPM) locking their group velocity in the cavity~\cite{WangOptica2017a}. These two traveling wavepackets only differ by their phase velocity, so that their extraction into the access waveguide results in two frequency comb components with the same comb tooth spacing yet with a shift of carrier-envelope offset (CEO) from one another. We refer to each component (\textit{i.e.} the auxiliary comb and the primary DKS) as \textit{colors}, defined by their respective phase velocity (CEO in the spectral domain). Interestingly, using this scheme enables a new phase matching condition at the second color~\cite{QureshiCommunPhys2022}, whose location is highly flexible since it can be widely tuned by hopping between cavity modes, or finely tuned through tuning within a cavity mode, without disrupting the DKS state. This results in the creation of DW(s) that can be harnessed, for instance, for spectral extension~\cite{MoilleNat.Commun.2021a,ZhangNatCommun2020}, where high power comb teeth can be generated at frequencies of interest. As these new dispersive waves are not solely governed by cavity dispersion, but are instead tunable based on the secondary laser frequency (\textit{i.e.} tuning the CEO difference between the colors), we refer to them as synthetic dispersive waves. Here, we harness both KIS and the synthetic dispersive waves produced in these multi-color solitons [\cref{fig:1}(c)]. By injecting an auxiliary laser on resonance with a cavity mode, hence shifted from its nearest comb tooth because of the resonator dispersion, we create a second color at the same group velocity as the DKS, yet with a spectral shift of tens of gigahertz from the DKS carrier-envelope offset frequency. We show that the second color, which automatically presents at least one DW, can undergo a power efficient KIS by a reference pump (a third laser injected into the resonator). Through XPM, the KIS of the second color is automatically transferred onto the DKS, enabling control and/or stabilization of the DKS repetition rate. Importantly, in this scheme the main pump does not require any locking, since all its noise will be absorbed by the spectral offset between the two comb components while the repetition rate remains locked through the OFD between the auxiliary and reference lasers. Also, as this color-KIS does not impose specific requirements on the resonator dispersion, it widely expands the range of devices and scenarios in which KIS can be utilized to synchronize DKS microcombs to optical references.


\subsection{Theoretical study of the color-KIS} 
The addition of other driving fields can be introduced in a single LLE accounting for their  phase and azimuthal offsets~\cite{TaheriEur.Phys.J.D2017}. Although this multi-pump LLE approach captures the Kerr-induced synchronization phenomenon~\cite{MoilleNature2023}, it can be challenging to study the different colors individually since they are all included in the single cavity field. Instead, one can obtain a system of nonlinearly coupled LLEs accounting for the DKS field $a_\mathrm{dks}$ and phase-offset second-color $a_\mathrm{sec}$, both driven by their respective pump $P_\mathrm{main}$ and $P_\mathrm{aux}$. Such formalism enables the re-normalization of their respective phase and allows study of each color independently (see Methods for the derivation):  

\begin{align}
    \frac{\partial a_\mathrm{dks}}{\partial t} &=%
     \left(-\frac{\kappa}{2} + i \delta\omega_\mathrm{main} \right)a_\mathrm{dks} %
     + i\Sigma_{\mu}D_\mathrm{int}(\mu)A_\mathrm{dks}\mathrm{e}^{i\mu\theta} \nonumber %
    \\&- i \gamma L(|a_\mathrm{dks}|^2 + 2|a_\mathrm{sec}|^2)a_\mathrm{dks} \label{eq:a_DKS}%
    \\&+ i\sqrt{\kappa_\mathrm{ext} P_\mathrm{main}}\nonumber\\
    \frac{\partial a_\mathrm{sec}}{\partial t} &= %
    \left(-\frac{\kappa}{2} + i \delta\omega_\mathrm{aux} - i D_\mathrm{int}(\mu_\mathrm{aux})\right)a_\mathrm{sec} \nonumber%
    \\&+  i\Sigma_{\mu}D_\mathrm{int}(\mu)A_\mathrm{sec}\mathrm{e}^{i\mu\theta}\nonumber %
    \\&- i \gamma L (2|a_\mathrm{dks}|^2 + |a_\mathrm{sec}|^2)a_\mathrm{sec} \label{eq:a_sec}%
    \\&+ i\sqrt{\kappa_\mathrm{ext} P_\mathrm{aux}}e^\mathrm{i\mu_{aux}\theta}\nonumber
\end{align}

\noindent with $A_\mathrm{dks}$ and $A_\mathrm{sec}$ the respective Fourier transforms of $a_\mathrm{dks}$ and $a_\mathrm{sec}$ in the azimuthal angle $\theta$/mode $\mu$ of the microring. The resonator properties that we are using can be found in the Methods.  The external parameters we experimentally control are the main and auxiliary pump powr $P_\mathrm{main} = 200$~mW and $P_\mathrm{aux} = 5$~mW, respectively, $\mu_\mathrm{aux}$ is the auxiliary pumped mode, and  $\delta\omega_\mathrm{main}$ and $\delta\omega_\mathrm{aux}$ are the detuning of the main and secondary pump relative to their respective pumped modes $\mu =0$ and $\mu = \mu_\mathrm{aux}$. Since the two colors are coupled through XPM, they exhibit the same group velocity but phase slip relative to one another, resulting in two equally-spaced comb components offset from one another by $\varpi = \delta \omega_\mathrm{aux}-\delta\omega_\mathrm{main} + D_\mathrm{int}(\mu_\mathrm{aux})$. As presented in refs~\cite{MoillearXiv2023, ShandilyaFront.Opt.2023}, this multi-color LLE formalism allows for straightforward understanding of the DW phase matching conditions, since any phase matching with the second color will occur at the zero crossing of $D_\mathrm{int}(\mu)  + \delta\omega_\mathrm{aux}$, allowing for the creation of a synthetic DW.  
For the geometry we study in the experiments presented later, higher-order dispersion is present, resulting in a roll-off in $D_\mathrm{int}$ with an eventual zero-crossing and generation of a DKS-DW.  We auxiliary pump away from the DKS-DW, with the higher-order dispersion providing a phase-matching point that is on the same side of the main pump [\cref{fig:2}(a)].

We simulate the multi-color LLE system with a main pump at $\mu=0$, an auxiliary pump at $\mu_\mathrm{aux} = -90$, and dispersion profile presented in \cref{fig:2}(a). We introduce the reference laser that triggers the synchronization through the inclusion in \cref{eq:a_sec} of an additional driving force $\sqrt{\kappa_\mathrm{ext}P_\mathrm{ref}}\mathrm{exp}\left(i\Omega t + i\mu_\mathrm{ref}\theta\right)$, with $\Omega =  \delta\omega_\mathrm{ref} - \delta\omega_\mathrm{main} + D_\mathrm{int}(\mu_\mathrm{ref}) - \varpi$ the frequency offset between the reference and the closest secondary comb tooth, with $\delta\omega_\mathrm{ref}$ the reference pump detuning, and $\mu_\mathrm{ref} =-45$  being the mode (relative to the main pump) at which we send the reference pump. We adiabatically and linearly sweep the detuning of the reference pump while seeding the initial solution with the soliton and secondary color solution for $a_\mathrm{dks}(\theta, t=0)$ and $a_\mathrm{sec}(\theta, t=0)$, respectively, and keeping both main and auxiliary pump fixed. 

When out-of-synchronization, the simulation highlights that the auxiliary and DKS colors travel at the same group velocity [\cref{fig:2}(c-d)] since the DKS drifts in $\theta$ at the same rate as the secondary color. Once the reference laser is within the KIS bandwidth, it captures the secondary-color comb tooth, stretching or compressing the microcomb and changing the secondary-color repetition rate [\cref{fig:2}(b)] with a slope given by the optical frequency division (OFD) factor set by $\Delta\mu = \mu_\mathrm{ref} -\mu_\mathrm{aux} = 45$, and hence controlling its group velocity. Because of the XPM binding, the DKS must follow any variation of group velocity imprinted onto the secondary color, which is apparent using the two-wave LLE formalism as they drift together. Hence, even if the the synchronization is performed on the secondary-color, the DKS repetition rate can be controlled. We can compare this scheme against a direct KIS at the same mode $\mu_\mathrm{ref}$, where the DKS comb does not present any DW and its shape closely follows a $\mathrm{sech}^2$ spectral envelope. Indeed, the KIS bandwidth $\Delta\omega_\mathrm{kis} = 2\omega_\mathrm{rep}\left|\Delta \mu\right| E_\mathrm{kis}$, with $E_\mathrm{kis}=\sqrt{\frac{\kappa_\mathrm{ext}}{\kappa^2}P_\mathrm{ref}|A_X(\mu_s)|^2}/E_\mathrm{X}$, $E_X$ being the total energy with $X$ the soliton or secondary color, and is directly proportional to the energy of the synchronization mode. Hence, for a fair comparison, all the resonator parameters, such as $D_1$, $\kappa$ and $\kappa_\mathrm{ext}$, must remain the same. Additionally, we have chosen this particular configuration as it conveniently results in the same OFD factor $\Delta\mu$ for the direct and color-KIS, with only a change of sign changing the direction of the $\omega_\mathrm{rep}$ trend against reference pump detuning. In the direct-KIS scenario, the available comb tooth energy $|A_\mathrm{sec}(\mu_s)|^2$ is considerably lower, by about 17~dB compared to the synthetic DW. In addition, the intracavity reference pump does not experience resonance enhancement (and could be considered single-pass) due to the close to 40~GHz offset of the comb and its nearest cavity mode -- with assumed losses resulting in an offset of more than 40 resonance linewidths -- preventing any significant resonant power build-up. This translates to a KIS bandwidth about 10 times smaller than the color-KIS bandwidth, with 130~MHz and 1.53~GHz synchronization bandwidths, respectively [\cref{fig:2}(b)].

\subsection{Experimental nonlinear characterization of the microring} 
Experimentally, we use a silicon nitride (\ce{Si3N4}) microring with a nominal thickness of $H=670$~nm embedded in \ce{SiO2} and a designed ring width of $RW=890$~nm, presenting a similar dispersion profile~[\cref{fig:3}a] as the one used in simulations in \cref{fig:2}. The DKS is obtained by pumping the ring at 283~THz with about 150~mW on-chip power, while the introduction of an auxiliary pump of about 2~mW in the 230~THz or 190~THz band only creates a secondary color without disturbing the DKS comb~[\cref{fig:3}b].  It is apparent that no direct-KIS could occur between the DKS and the secondary color since the two comb components are separated by about 40~GHz (resolved with an optical spectrum analyzer; see left inset to~\cref{fig:3}b). We further verify the flexibility and efficiency of the creation of new synthetic DWs by auxiliary pumping different modes in the 190~THz band and 230~THz band. Figure~\ref{fig:3}c-d highlights the creation of a new DW in these other laser bands, and how the synthetic DW spectral position, which  will determine where the KIS reference laser is applied, can be flexibly controlled by tuning the auxiliary laser frequency. A significant improvement of 25~dB (190~THz band auxiliary pump) or 32~dB (230~THz band auxiliary pump) of the comb tooth power available at the new DW where the color-KIS will happen is observed compared to the nearest DKS comb tooth. Along with the possibility to send the reference on-resonance, it confirms the theoretical investigation previously conducted that color-KIS will be more efficient than conducting a direct-KIS at these modes, since the color-KIS energy will be about 30$\times$ larger than the one for direct KIS, and we were not able to observe direct KIS at these modes of interest between $\mu = -51$ to $\mu = -60$.


\subsection{DKS repetition rate entrainment from color-KIS}
To characterize the color-KIS we measure at the same time the repetition rate $\omega_\mathrm{rep}$ and the frequency offset between the reference laser and the secondary color $\Omega$ (see experimental setup in the Supplementary Information S.1). As our focus is on DKS repetition rate control, the repetition rate of the system is measured only at the DKS color. This is done by measurement around 272~THz, where the secondary color presents negligible power in comparison to the DKS, as evident through a comparison of the spectra with and without the auxiliary/reference pumps [\cref{fig:3}b].

First, we use the 190~THz CTL as the auxiliary pump generating the secondary color, and keep its detuning fixed. We use the 230~THz CTL to act as the reference and it is frequency swept close to the secondary color comb tooth [\cref{fig:4}a]. We use the different modes that have been presented in \cref{fig:3}c-d, resulting in a variation of $\Delta\mu=25$ to $\Delta\mu=41$. A clear change in the repetition rate following a linear trend with the reference detuning occurs when $\Omega$ is locked to zero, consistent with conventional KIS behavior, except in these measurements $\Omega=0$ corresponds to a phase locking at the second color while the measurement of the repetition rate variation $\Delta \omega_\mathrm{rep}$ is at the DKS color, highlighting the XPM interplay linking the color-KIS to the DKS. We verify the linear trend of the repetition rate entrainment by overlaying the reference laser detuning divided by $\Delta\mu$, which closely matches $\Delta \omega_\mathrm{rep}$. We note here that this linear trend is plotted without offset or scaling, since the zero detuning is accurately calibrated using the technique presented in the Supplementary Information S.2 and since the OFD factor is a simple comb tooth counting. Additionally, we note a discrepancy in KIS bandwidth between the simulation presented in~\cref{fig:2} and the experimental result. This is likely a result of resonator-waveguide coupling dispersion that leads to a non trivial variation of $\kappa_\mathrm{ext}$ and $E_\mathrm{sec}$, and which is not included in the LLE model. We can then switch the system, where the 230~THz CTL becomes the fixed auxiliary pump creating the secondary color while the 190~THz CTL becomes the swept reference pump [\cref{fig:4}b]. Similar results as previously described are obtained, yet this time the OFD coefficient $\Delta\mu = \mu_\mathrm{ref} - \mu_\mathrm{aux}$ is negative. The repetition rate entrainement slope becomes negative when $\Omega =0$, and is once again accurately captured by a simple linear trend accounting for $1/\Delta\mu$. We note that this experiment has been conducted while maintaining the same exact DKS state, highlighting the little impact the secondary color and its synchronization has on the stability of the DKS, further providing flexibility to the system for on-the-fly reconfigurable OFD. 

\subsection{DKS repetition rate phase noise from color-KIS}

We investigate the phase noise of the DKS repetition rate while the second color becomes synchronized. For the different OFD factors presented in~\cref{fig:4}, we enter the color-KIS regime and lock both auxiliary and reference lasers to independent fiber Mach-Zehnder interferometers (MZIs) with a 10~MHz free spectral range (FSR), while the main pump and cooler pump are free running. Using a third similar MZI with a 40~MHz FSR, we measure the optical phase noise of both pump lasers. We note that in this measurement we account for the noise of both lasers; however, since the 230~THz laser carries larger phase noise while locked than the 190~THz laser (see Supplementary Information S.3), only accounting for the former to compare against $\omega_\mathrm{rep}$ phase noise enables verification of OFD. For each value of $\Delta\mu$, the repetition rate is well-matched to the laser phase noise divided by the OFD factor $\Delta \mu^2$ [\cref{fig:5}]. Although the repetition rate entrainement presented in~\cref{fig:4} already demonstrated OFD based on the color-KIS technique, these additional phase noise measurements highlight that the XPM-mediated transfer of synchronization of the second color onto the DKS does not carry additional noise, so that color-KIS can be used to stabilize a DKS repetition rate indirectly. We further note that for $\Delta\mu = 30$, we have swept the reference pump over a range of about 120~MHz at a 75~Hz rate following a triangular waveform, and confirmed that no additional noise is present in the repetition rate phase noise. This is consistent with $\omega_\mathrm{rep}$ being locked for both colors, while the frequency offset between the two combs is not. Essentially, while in the color-KIS regime, all of the phase noise of the main pump is transferred onto the carrier-envelope offset noise of the DKS.

\begin{figure}
    \begin{center}
        \includegraphics{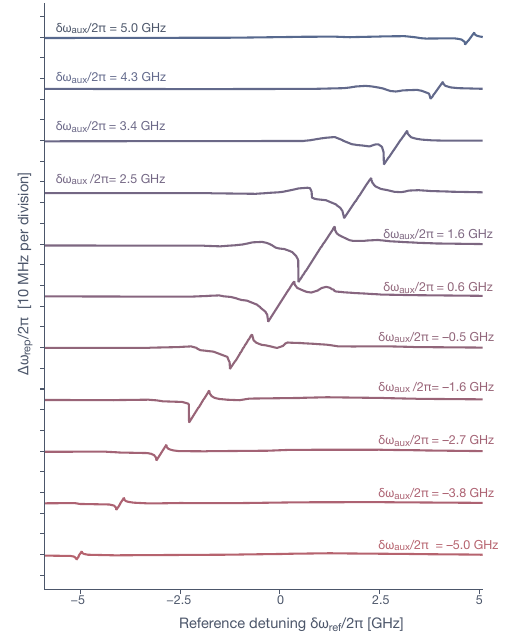}
        \caption{\label{fig:6} \textbf{Tunability of the secondary color comb tooth for color-KIS frequency versatility -- }%
        Using the same DKS previously presented, we sweep the reference laser around 226~THz while the auxiliary pump is parked around the cavity mode at 192~THz, leading to an OFD factor $\Delta\mu=34$ consistent with the $\Delta\omega_\mathrm{rep}$ entrainment slope against the reference detuning $\delta\omega_\mathrm{ref}$. The auxiliary pump detuning $\delta\omega_\mathrm{aux}/2\pi$ is tuned in steps over 10~GHz and the color-KIS frequency window is tuned accordingly, demonstrating the possibility to match a precise target frequency at which the synchronization shall happen. 
        }
    \end{center}
\end{figure}

\subsection{Large tunability of color-KIS frequency}
Finally, we address the large tunability that the color-KIS enables. In contrast to the standard-KIS scheme, the color-KIS does not occur on the fixed frequency markers defined by the repetition rate and the main pump (i.e., the CEO frequency). Instead, the secondary colors live at a CEO offset from the main pump, which can be widely tuned through mode hopping between cavity resonances, as previously demonstrated~[Fig.~\ref{fig:3}(c)-(d)]. Additionally, a finer tuning of the CEO frequency offset of the second color can be obtained through auxiliary pump detuning from its resonance, resulting in the tuning of the synthetic DW~\cite{QureshiCommunPhys2022}. In particular, since the quality factor and coupling regime requirements are much less stringent than for DKS generation, one can harness the natural overcoupling at lower frequencies than the main pump from straight-waveguide coupling~\cite{MoilleOpt.Lett.2019b}, resulting in large secondary color CEO offset tunability while still dropping significant auxiliary power from the waveguide to the cavity. The repetition rate of the DKS remaining mostly unperturbed with the auxiliary pump detuning enables one to tune the secondary-color comb tooth at which the color-KIS occurs. We demonstrate this effect experimentally using a OFD factor $\Delta\mu=34$ between the auxiliary and reference pump in the same DKS presented previously [\cref{fig:6}]. The reference laser is swept, similarly to~\cref{fig:4}, while the auxiliary pump is fixed. The detuning of the auxiliary pump $\delta\omega_\mathrm{aux}$ enables all the comb teeth of the secondary color to also be shifted by $\delta\omega_\mathrm{aux}$, thus shifting the color-KIS window accordingly. We demonstrate a tuning of over 10~GHz of the KIS window with $\approx$5~mW of on-chip power for both pumps, respectively, corresponding to a tunability of about ten times the optimized color-KIS bandwidth at this power when both pumps are at the center of their respective resonances. This tunability is only limited by the available power of the auxiliary pump, and could be further increased by using a more powerful laser while remaining in the thermal stability range of the DKS. Such tunability is a key feature of the color-KIS scheme, as it enables synchronization to occur at a precise target frequency, for instance to match a fixed frequency ultra-stable laser.


\subsection{Discussion}
In conclusion, we have demonstrated a novel way to efficiently access the Kerr-induced synchronization regime, enabling cavity soliton repetition rate control, and optical frequency division with all-optical synchronization of the comb. Using the fundamental group velocity binding between a DKS color and a secondary color created by an auxiliary laser traveling in the same microring resonator, we create new phase-matching conditions which enable new synthetic DWs that can be used for synchronization, and whose spectral locations can be flexibly tuned. Compared to synchronization to a reference laser at the same frequency using a single-pumped DKS, the comb tooth power is orders of magnitude larger while providing on-resonance operation for the reference laser to efficiently transfer its power inside the resonator. We show theoretically that the KIS bandwidth for a similar situation between direct and color-KIS is increased by at least an order of magnitude. Experimentally, we have demonstrated this effect using an integrated microring resonator, where repetition rate entrainment according to the optical frequency division factor between the auxiliary and the reference pump is verified. The spectral separation between the auxiliary and reference pump, and hence the division factor, can be widely controlled with respect to the underlying cavity dispersion without disrupting the stability of the DKS. We also show that the locking of the group velocity of the second-color through auxiliary and reference pump locking in the color-KIS regime is transferred onto the DKS repetition rate in a noiseless fashion, despite the indirect control of the DKS through the XPM-coupled secondary color. Finally, we have demonstrated the versatility of the color-KIS approach in comparison to the standard KIS one, where the synchronization frequency (and window) can be largely tuned without impacting the DKS. Our work enables a new efficient way to control the large repetition rate of integrated microresonator combs, in particular in situations where the cavity dispersion could otherwise prevent direct KIS.

We further discuss a few important implications of the color-KIS effect. For example, KIS-enabled optical frequency division has recently been shown in the context of low noise microwave generation with a greatly simplified architecture~\cite{Sun2024}. Our color-KIS scheme enables further important simplifications. Indeed, it is interesting to note that the main pump does not play a role in the repetition rate control and stabilization since the CEO offset between colors $\varpi$ is unlocked and therefore uncoupled from optical frequency division approach. This enables the main pump to be freely tuned as needed to generate the DKS, without further conditions on its stability. In addition, although we have mostly focused on demonstrating the control of the single repetition rate of the different colors, it is important to point out that the secondary color forms its own frequency comb whose repetition rate can also be detected and stabilized. As a result, the color-KIS scheme brings the possibility to coherently phase lock an auxiliary laser to an ultra-narrow linewidth laser, through the repetition rate feedback onto the auxiliary laser. As noted above, in color-KIS, such a scheme is independent of the main pump; indeed, the main DKS essentially acts as an intermediary whose XPM interactions with the second color enable a single group velocity to exist. We note that while thus far, color-KIS has required an additional laser in comparison to direct KIS, one could leverage the cooler pump to also act as the auxiliary pump that creates the synthetic DW. This simplification of the color-KIS architecture is still effective when the auxiliary pump is counter-propagating, similar to the synchronization of two counter propagating solitons~\cite{YangNaturePhoton2017}, and is demonstrated in Supplementary Information Section S4. 
Finally, the creation of synthetic DWs can be extended to other colors, which has been demonstrated to expand the comb bandwidth well beyond an octave~\cite{MoilleNat.Commun.2021a}, and hence the color-KIS scheme may find application in the stabilization of such ultra-broadband integrated frequency combs. Beyond the above, our work could find direct application in distance ranging while enabling new techniques for fundamental studies of DKS dynamics in micrometer-scale resonators.

                       
\section{Acknowledgments}
\noindent The Scientific colour map batlow~\cite{Crameri2023} and subsequent color set is used in this study to prevent visual distortion of the data and exclusion of readers with colour-vision deficiencies~\cite{CrameriNatCommun2020}. 

\noindent We acknowledge partial funding support from the Space Vehicles Directorate of the Air Force Research Laboratory, the Atomic–Photonic Integration programme of the Defense Advanced Research Projects Agency, and the NIST-on-a-chip program of the National Institute of Standards and Technology. P.S and C.M. acknowledges support from the Air Force Office of Scientific Research (Grant No. FA9550-20-1-0357) and the National Science Foundation (Grant No. ECCS-18-07272). We thank Sean Krzyzewski and Marcelo Davan\c{c}o for insightful feedback.

\noindent G.M dedicates this work to T.B.M -- always.

\section{Author Contributions}
G.M. and K.S. led the project. G.M. designed the resonators and performed the measurements and simulations. P.S. and C.M. helped with the theoretical and numerical understanding. A.N. and G.C. helped with the experimental understanding. K.S. helped with data analysis. G.M. and K.S. wrote the manuscript, with input from all authors. All the authors contributed and discussed the content of this manuscript.

\section{Competing Interests}
G.M., C.M. and K.S have submitted a provisional patent application based on aspects of the work presented in this paper. The remaining authors declare no competing interests.

\clearpage

                                 
\section{Methods}
\subsection{Derivation of the coupled LLE formalism}
We start with the multi-pumped LLE (MLLE) with two driving forces as presented in ref~\cite{TaheriEur.Phys.J.D2017}
\begin{equation}
    \begin{split}
    \label{eq:method_mLLE} 
    \frac{\partial a}{\partial t} &=\left(-\frac{\kappa}{2} + i\delta \omega_{\mathrm{main}}\right)a \\
        &+ i\sum_\mu D_\mathrm{int}(\mu)A\mathrm{e}^{i\mu\theta} - i\gamma L |a|^2 a \\
        & + i\sqrt{\kappa_\mathrm{ext}P_{\mathrm{main}}} \\
        & + i\sqrt{\kappa_\mathrm{ext}P_{\mathrm{aux}}}\mathrm{e}^{i(\delta \omega_\mathrm{main}-\delta\omega_\mathrm{aux} + D_\mathrm{int}(\mu_\mathrm{aux}))t + i\mu_\mathrm{aux}\theta}
    \end{split}
\end{equation}
with $A = \mathrm{FT}[a]$ the azimuthal Fourier transform of $a$, $\delta\omega_\mathrm{main}$ and $\delta\omega_\mathrm{aux}$ the detuning of the main and auxiliary pump relative to their mode $\mu = 0$ and $\mu = \mu_\mathrm{aux}$, respectively, and $D_\mathrm{int}(\mu_\mathrm{aux})$ is the integrated dispersion, computed with respect to the main pump, for the auxiliary pumped mode $\mu_\mathrm{aux}$. Here, we have defined the detuning of a pump `$p$' with frequency $\omega_p$ from its nearest cavity resonance frequency $\omega(\mu=\mu_p)$ as $\delta\omega_p = \omega(\mu=\mu_p) - \omega_p$. In this notation, $\delta\omega_p>0$ correspons to a red-detuned pump. The integrated dispersion is defined as $D_\mathrm{int}(\mu) = \omega_\mathrm{res}(\mu) - \omega_\mathrm{res}(\mu=0) - D_1 \mu_n$, where $D_1$ is the cavity free spectral range around the main pump resonance $\omega_\mathrm{res}(\mu=0)$. In this formalism of the MLLE, we have assumed that the soliton rotates in the $-\theta$ direction.

We assume that two colors form in the cavity, the second color experiencing a phase offset $\varpi t = \left[\delta \omega_\mathrm{main}-\delta\omega_\mathrm{aux} + D_\mathrm{int}(\mu_\mathrm{aux})\right]t$ from the soliton. Therefore $a$ can be written as,
\begin{equation}
    \label{eq:method_colorsplit}
    a = a_\mathrm{dks} + a_\mathrm{sec}\mathrm{e}^{i\varpi t},
\end{equation}

Substituting~\cref{eq:method_colorsplit} in~\cref{eq:method_mLLE} and retaining only the phase-matched terms,
\begin{equation}
    \begin{split}
        &\frac{\partial a_\mathrm{dks}}{\partial t} + \frac{\partial a_\mathrm{sec}}{\partial t}\mathrm{e}^{i\varpi t} + i\varpi a_\mathrm{sec}\mathrm{e}^{i\varpi t} = \\
        &\left(-\frac{\kappa}{2} + i\delta \omega_{\mathrm{main}} \right)\left(a_\mathrm{dks} +  a_\mathrm{sec}\mathrm{e}^{i\varpi t}\right)\\
        & + i\sum_\mu D_\mathrm{int}(\mu)A_\mathrm{dks}\mathrm{e}^{i\mu\theta} \\
        &+ i\sum_\mu D_\mathrm{int}(\mu)A_\mathrm{sec}(\mu)\mathrm{e}^{i\mu\theta}\mathrm{e}^{i\varpi t} \\
        &-i\gamma L \left(|a_\mathrm{dks}|^2 + 2|a_\mathrm{sec}|^2\right)a_\mathrm{dks} \\
        &- i\gamma L \left(|a_\mathrm{sec}|^2 + 2|a_\mathrm{dks}|^2\right)a_\mathrm{sec}\mathrm{e}^{i\varpi t} \\
&+ i\sqrt{\kappa_\mathrm{ext}P_\mathrm{main}}+i\sqrt{\kappa_\mathrm{ext}P_{\mathrm{aux}}}\mathrm{e}^{(i\varpi t + i\mu_\mathrm{aux}\theta)}
    \end{split}
\end{equation}
Separating the terms with and without $\mathrm{e}^{i\varpi t}$, we get the equation for $a_\mathrm{dks}$,
\begin{equation}
    \begin{split}
        \frac{\partial a_\mathrm{dks}}{\partial t} &=\left( -\frac{\kappa}{2} + i\delta\omega_\mathrm{main}\right)a_\mathrm{dks} \\
        &+ i\sum_\mu D_\mathrm{int}(\mu)a_\mathrm{dks}\mathrm{e}^{i\mu\theta} \\
        &- i\gamma L (|a_\mathrm{dks}|^2 + 2|a_\mathrm{sec}|^2)a_\mathrm{dks} \\
        &+ i\sqrt{\kappa_\mathrm{ext}P_\mathrm{main}},
    \end{split}
\end{equation}
which is equation (1) in the main text, and for $a_\mathrm{sec}$,

\begin{equation}
    \begin{split}
        \frac{\partial a_{\mathrm{sec}}}{\partial t} &+ i\varpi a_{\mathrm{sec}}= \left(- \frac{\kappa}{2} + i\delta\omega_{\mathrm{main}}\right)a_{\mathrm{sec}} \\
        &+ i\sum_{\mu'} D_{\mathrm{int}}(\mu')a_{\mathrm{sec}}(\mu')\mathrm{e}^{i\mu\theta} \\ &- i\gamma L (|a_{\mathrm{sec}}|^2 + 2|a_{\mathrm{dks}}|^2)a_{\mathrm{sec}} \\
        &+ i\sqrt{\kappa_{\mathrm{ext}}P_{\mathrm{sec}}}\mathrm{e}^{i\mu_\mathrm{aux}\theta},
    \end{split}
\end{equation}
Using the definition of $\varpi = \delta \omega_{\mathrm{main}}-\delta\omega_{\mathrm{aux}} + D_{\mathrm{int}}(\mu_{\mathrm{aux}})$ gives,
\begin{equation}
    \begin{split}
        \frac{\partial a_\mathrm{sec}}{\partial t} &= \left( -\frac{\kappa}{2} + i\delta\omega_\mathrm{aux} - iD_\mathrm{int}(\mu_\mathrm{aux})\right)a_\mathrm{sec} \\
        &+ i\sum_{\mu} D_\mathrm{int}(\mu)a_\mathrm{sec}(\mu)\mathrm{e}^{i\mu\theta} \\ 
        &- i\gamma L \left(|a_\mathrm{sec}|^2 + 2|a_{dks}|^2 \right)a_\mathrm{sec} \\
        &+ i\sqrt{\kappa_\mathrm{ext}P_\mathrm{aux}}\mathrm{e}^{i\mu_\mathrm{aux}\theta},
    \end{split}
\end{equation}

\vspace{1em}
Now, we can add a reference pump intended to synchronize the second color to the MLLE as, 
\begin{equation}
\begin{split} 
    \frac{\partial a}{\partial t} &=  \left(-\frac{\kappa}{2} + i\delta \omega_\mathrm{main}\right)a \\
     & + i\sum_\mu D_\mathrm{int}(\mu)A\mathrm{e}^{i\mu\theta} - i\gamma L |a|^2 a
    + i\sqrt{\kappa_{ext}P_{\mathrm{main}}} \\&+ i\sqrt{\kappa_{ext}P_{\mathrm{aux}}}\mathrm{e}^{i(\delta \omega_\mathrm{main}-\delta\omega_\mathrm{aux} + D_\mathrm{int}(\mu_\mathrm{aux}))t + i\mu_\mathrm{aux}\theta} \\
    &+ i\sqrt{\kappa_\mathrm{ext}P_\mathrm{ref}}\mathrm{e}^{i(\delta\omega_\mathrm{main} - \delta\omega_\mathrm{ref} + D_\mathrm{int}(\mu_\mathrm{ref}))t + i\mu_\mathrm{ref}\theta}
    \end{split}
\end{equation}
Substituting the solution and separating the equations like before while grouping the reference pump with the equation for the second color, the coupled LLE for $a_\mathrm{dks}$ remains unchanged, while $a_\mathrm{sec}$ grouped with the reference pump becomes,
\begin{equation}
    \begin{split}
        \frac{\partial a_\mathrm{sec}}{\partial t} &= \left(- \frac{\kappa}{2} + i\delta\omega_\mathrm{aux} - iD_\mathrm{int}(\mu_\mathrm{aux})\right)a_\mathrm{sec} \\
        & + i\sum_{\mu} D_\mathrm{int}(\mu)A_\mathrm{sec}(\mu)\mathrm{e}^{i\mu\theta} \\ 
        &- i\gamma L \left(|a_\mathrm{sec}|^2 + 2|a_\mathrm{dks}|^2 \right)a_\mathrm{sec} \\
        & + i\sqrt{\kappa_\mathrm{ext}P_\mathrm{aux}}\mathrm{e}^{i\mu_\mathrm{aux}\theta} \\
        &+ i\sqrt{\kappa_\mathrm{ext}P_\mathrm{ref}}\mathrm{e}^{i[\delta\omega_\mathrm{aux}-\delta\omega_\mathrm{ref} + D_\mathrm{int}(\mu_\mathrm{ref}) - D_\mathrm{int}(\mu_\mathrm{aux})]t + i\mu_\mathrm{ref}\theta},
    \end{split}
\end{equation}
With the definition of $\varpi  = \delta \omega_\mathrm{main}-\delta\omega_\mathrm{aux} + D_\mathrm{int}(\mu_\mathrm{aux})$ as the offset of the secondary color field relative to the DKS, one can define $\Omega =  \delta\omega_\mathrm{main} - \delta\omega_\mathrm{ref} + D_\mathrm{int}(\mu_\mathrm{ref}) - \varpi$ as the offset of the reference from its nearest secondary comb tooth. We obtain,
\begin{equation}
    \begin{split}
        \frac{\partial a_{sec}}{\partial t} &= \left(- \frac{\kappa}{2} + i\delta\omega_\mathrm{aux} - iD_\mathrm{int}(\mu_\mathrm{sec})\right)a_\mathrm{sec} \\
        &+ i\sum_{\mu} D_\mathrm{int}(\mu)A_\mathrm{sec}(\mu)\mathrm{e}^{i\mu\theta} \\
        &- i\gamma L \left(|a_\mathrm{sec}|^2 + 2|a_\mathrm{dks}|^2 \right)a_\mathrm{sec} \\
        &+ i\sqrt{\kappa_\mathrm{ext}P_\mathrm{sec}}\mathrm{e}^{i\mu_\mathrm{aux}\theta} \\
        &+ i\sqrt{\kappa_\mathrm{ext}P_\mathrm{ref}}\mathrm{e}^{(i\Omega t + i\mu_\mathrm{ref}\theta)},
    \end{split}
\end{equation}
which is equation (2) in the main text.

\subsection{Simulation parameters}
In the equation~\cref{eq:a_DKS,eq:a_sec}, $\gamma = 2.3$~W\textsuperscript{-1}$\cdot$m\textsuperscript{-1} is the effective nonlinearity calculated at the main pump~\cite{MoilleNature2023,MoilleNat.Commun.2021a}, $\kappa/2\pi = 1.1$~GHz is the total loss rate, $\kappa_\mathrm{ext} = \frac{1}{2}\kappa$ is the external coupling rate (\textit{i.e.}, $Q_c=Q_i=500 \times 10^3$ are the coupling and intrinsic quality factor respectively), $L = 2\pi \times 23$~{\textmu}m is the resonator circumference, and $D_\mathrm{int}(\mu)$ is the integrated dispersion accounting for the DKS repetition rate $\omega_\mathrm{rep}/2\pi \approx 1$~THz.

\subsection{Resonator design}
The photonic chips were fabricated in a commercial foundry in the same fashion as the ones presented in~\cite{MoilleNature2023}. The microring resonator is made of silicon nitride (\ce{Si3N4}) embedded in silicon dioxide (\ce{SiO2}). Its dimensions consist of an external ring radius $RR=23$~{\textmu}m, \ce{Si3N4} thickness of $H=670$~nm, and ring width of $RW=890$~nm. We couple to the ring using a $460$~nm wide waveguide that is at a gap distance from the ring of $G=500$~nm. The light is injected through facet coupling with a lensed fiber into the inverse-taper waveguide that exhibits a $W=250$~nm width at the facet. The insertion losses from fiber to chip are approximately 2.3~dB.
\subsection{Comb generation}
The experimental setup is discussed in detail in Supplementary Information S.1. To generate the DKS, we use a 283~THz pump with about 150~mW on-chip power. The DKS is adiabatically accessed using a 308~THz counter-propagating cooler laser that pumps a fundamental transverse magnetic mode, which thermally stabilizes the resonator and enables long-term operation of the DKS state~\cite{ZhangOptica2019a, ZhouLightSciAppl2019}. To generate the secondary color, we use continuously tunable lasers (CTLs) spanning between 184~THz to 198~THz with about 2~mW of on-chip power, and 227~THz to 232~THz with about 1~mW of on-chip power. One of these lasers acts as the auxiliary pump that creates the second color (i.e., is fixed), while the other is used as the reference The higher-order dispersion results in $D_\mathrm{int}(\mu)$ roll-off, enabling the creation of new synthetic dispersive wave phase-matching in the 230~THz range (\textit{i.e.,} 1300 nm range) for auxiliary pumping in the 190~THz range (\textit{i.e.,} 1550~nm range), and vice-versa. The experimental dispersion is measured using wavemeter-calibrated resonance frequencies of the first order transverse electric mode [\cref{fig:3}a]. %

\subsection{Data acquisition method}
To measure the repetition rate, wee use an electro-optic comb apparatus~\cite{MoilleNature2023, StonePhys.Rev.Lett.2020} to modulate two adjacent DKS comb teeth and frequency translate $\omega_\mathrm{rep} / 2\pi\approx 1$~THz to a detectable microwave beat note; an example spectrum is shown in the right inset to Fig.~\ref{fig:3}b. We sweep the reference laser using an electrical ramp signal sent to the continuously tunable laser (CTL) piezo element, enabling small detuning to be studied, and measure both $\omega_\mathrm{rep}$ and $\Omega$ by recording simultaneously their temporal trace with a fast oscilloscope. We then process these data to create a spectrogram where the temporal dependence of $\omega_\mathrm{rep}$ and $\Omega$ are obtained. The detuning calibration is obtained through beating the reference laser against the secondary color comb tooth while bypassing the resonator (i.e., no nonlinearity, hence no synchronization), calibrating the amplitude and the zero of the detuning. For more details, the complete procedure is described in Supplementary Information S.2. 

\section{Data availability}
The data that supports the plots within this paper and other findings of this study are available from the corresponding authors upon request.

\section{\large Code availability}
The simulation code is available from the authors through the pyLLE package available online~\cite{MoilleJ.RES.NATL.INST.STAN.2019}, with a modification that is available upon reasonable request, using the inputs and parameters presented in this work.


\clearpage
\onecolumngrid
\renewcommand{\appendixpagename}{\large\centering Supplementary Information: \mytitle}
\appendix   
\appendixpage
\renewcommand{\thesection}{S.\arabic{section}}
\renewcommand\thefigure{S.\arabic{figure}}    
\setcounter{figure}{0}    

\section{Experimental Setup
\label{supsec:exp_setup}}
\begin{figure*}[h]
    \centering
    \includegraphics{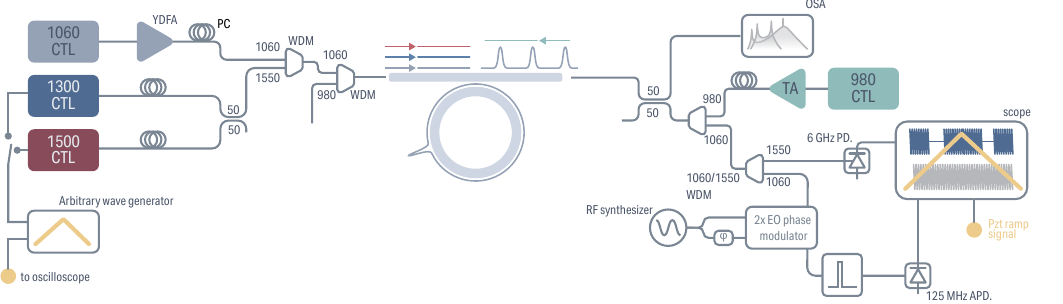}
    \caption{\label{figsup:setup}
    \textbf{Experimental setup}. A main pump around 283~THz is obtained from an amplified 1060~nm CTL providing about 150~mW of on-chip power to generate a DKS, which is then adiabatically accessed thanks to the amplified and counterpropagating 980~nm CTL cooler pump that thermally stabilizes the resonator. We introduce an auxiliary pump and a reference pump to create the secondary color and trigger the color-KIS with 1300~nm and 1550~nm CTLs. We measure the DKS repetition rate using two cascaded EO phase modulators which we apply to two adjacent Kerr comb lines around 1100~nm. EO-generated sidebands of the two DKS teeth span across the microcomb repetition rate, with high-order sidebands that come close to merging. We filter around this frequency, allowing retrieval of the beat note with a slow ($<$50~MHz bandwidth) APD. The beating between the secondary color and the reference pump is obtained with negligible DKS power thanks to the WDM. Both the repetition rate and the reference beat note are measured in the time domain and then processed and calibrated to obtain the data presented in this work. Other elements such as power meters and photodiodes to verify transmission levels are not displayed to avoid over-complicating the schematic. %
    CTL: continuously tunable laser, YDFA: ytterbium-doped fiber amplifier, TA: tapered amplifier, PC: polarization controller, WDM: wavelength demultiplexer, RF: radiofrequency, OSA: optical spectrum analyzer, PD: photodiode, APD: avalanche photodiode, EO: electro-optic %
    }
\end{figure*}

The experimental setup used in this work is depicted in \cref{figsup:setup}. We use a 1060~nm continuously tunable laser (CTL) amplified with an ytterbium-doped fiber amplifier (YDFA) to generate enough power at 283~THz to pump the microring to harness the resonator's third-order nonlinearity. An on-chip power of about 150~mW is sufficient to reach a dissipative Kerr soliton (DKS) state in the fundamental transverse electric mode (TE) of the 23~{\textmu}m \ce{Si3N4} microring resonator. To help access this state, we use a 980~nm CTL amplified using a taper amplifier (TA) and set to be counterpropagative and cross-polarized relative to the main pump, so that this cooler pump only thermally stabilizes the resonator for adiabatic access to the DKS~\cite{ZhangOptica2019a, ZhouLightSciAppl2019} while minimizing potential nonlinear mixing with the main pump and/or the DKS. We use a combinations of wavelength demultiplexers (WDMs) at the input and output to avoid any injection of the cooler (main pump) into the main pump (cooler). The auxiliary and reference laser from the 1300~nm and 1550~nm CTLs are injected in the resonator by combining them with the main pump using a WDM, and are also set in the TE polarization. We tap 50~\% of the output to probe the microcomb optical spectrum using an optical spectrum analyzer (OSA). We use another WDM at the other 50~\% output to separate wavelengths close to 1100~nm, which are electro-optically (EO) modulated at $\Omega_\mathrm{RF}=17.87892$~GHz. The EO-generated sidebands of two adjacent DKS comb teeth  fill the DKS repetition rate over $N=56$ EOcomb teeth, providing a measure of the DKS repetition rate $\omega_\mathrm{rep} = N \times \Omega_\mathrm{RF} - \delta\omega_\mathrm{rep} = 1001.21952$~GHz $-~\delta\omega_\mathrm{beat}$ (the minus sign is configuration dependent and has been verified for our experiments), with $\delta\omega_\mathrm{beat}$ the EOcomb beat note that can be detected by a 50~MHz avalanche photodiode (APD), similar to the methods presented in~\cite{StonePhys.Rev.Lett.2020,MoilleNature2023}. The other WDM output is used to measure the beat note {$\Omega$} between the secondary color and the reference pump. Both signals are processed in the time domain using a high speed oscilloscope. We trigger the scope with the ramp signal used to sweep the piezo element of the CTL, obtaining a temporal trace which can be processed and converted into a spectrogram, enabling instantaneous frequency measurement. The calibration of this technique is further discussed in the next section.

\section{Repetition rate and detuning calibrations from temporal traces
\label{supsec:det_cal}}

\begin{figure*}[h]
    \centering
    \includegraphics{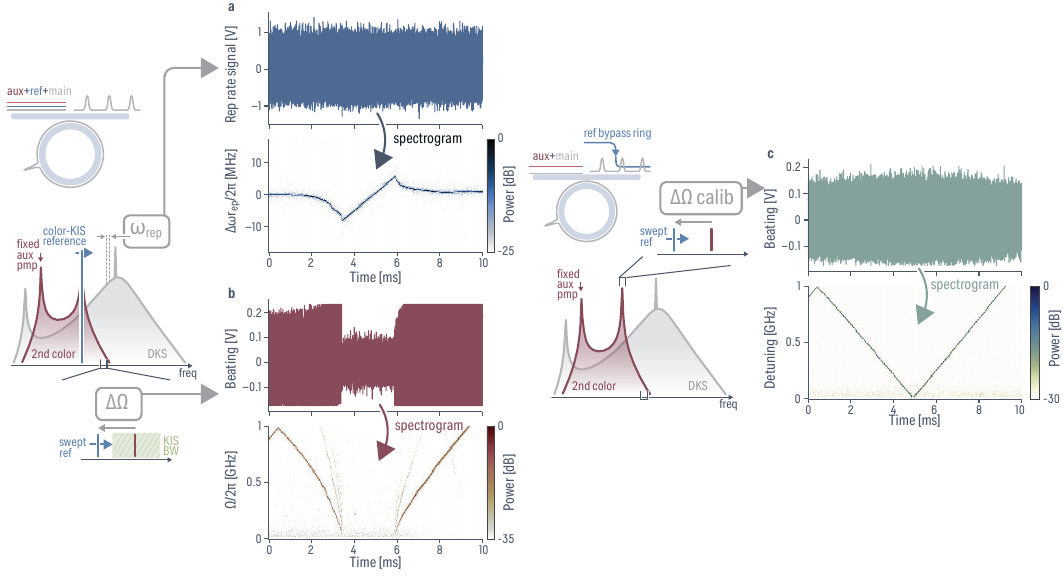}
    \caption{\label{figsup:calib}
    \textbf{Repetition rate, detuning beat note and calibration protocols}. %
    \textbf{a-b} Simultaneous measurement of the repetition rate $\omega_\mathrm{rep}$ and the beat note between the reference and the second color $\Omega$. We record the temporal trace while sweeping the reference laser with a triangular waveform that actuates the CTL piezo element. The direction of the sweep is known from the CTL characteristics: high-to-low voltage results in high-to-low reference laser frequency, and the oscilloscope is triggered on the raising slope of the triangular waveform. Applying a consecutive fast Fourier transform (through the \textit{scipy} python package~\cite{VirtanenNatMethods2020}) allows us to retrieve the instantaneous frequency of the signal. %
    \textbf{c} Calibration of the detuning to convert the temporal frequency trace from the spectrogram maxima in \textbf{a} and \textbf{b}, performed for each $\Delta\mu$ presented in \cref{fig:4}. We bypass the resonator with the reference laser and beat it directly against the second color comb tooth. We record the temporal trace and apply a fast Fourier transform to retrieve the instantaneous frequency. In this calibration nonlinear effects associated with the reference laser cannot occur, and the detuning against the comb tooth can be directly obtained, including the zero reference for the detuning. The obtained frequency beat is flipped if needed according to knowledge if the detuning is positive or negative based on the triangular waveform applied. 
    }
\end{figure*}

The EOcomb beat note $\delta \omega_\mathrm{beat}$, which directly lets us precisely characterize the repetition rate $\omega_\mathrm{rep}$, and the reference pump and secondary-color beat $\Omega$ are measured in the optical domain using a fast oscilloscope~[\cref{figsup:calib}a-b], with a temporal resolution of 320~ps. The piezo element is swept at a 50~Hz rate around the detuning of interest. We record simultaneously the temporal trace of $\delta \omega_\mathrm{beat}$ and $\Omega$, which we process using the \textrm{scipy.signal.spectrogram} python package~\cite{VirtanenNatMethods2020}. We use a binning slice of $30\times10^3$ points, allowing a frequency resolution of about 104~kHz. The fast Fourier transform (FFT) windowing is performed using a Hanning filter, with an overlap of $10\times10^3$ points. We obtain a spectrogram with a clear peak for each time sample that correspond to either $\delta \omega_\mathrm{beat}$, from which we extract $\omega_\mathrm{rep}$ using the RF synthesizer frequency and the number of comb teeth as previously described, or $\Omega$. 

However, the detuning needs to be calibrated accurately. We harness the fact that the frequency comb, if unsynchronized, can act as a local oscillator. To this extent, for each of the different $\Delta\mu$ measurements shown in~\cref{fig:4}, we proceed to calibrate the detuning by bypassing the microresonator with the reference and directly beating it with the closest secondary color comb tooth (\textit{i.e.}, no nonlinearity present since we bypass the ring, and hence no KIS) as shown in~\cref{figsup:calib}c. Once again, we proceed to obtain the spectrogram from the temporal trace, from which we can calibrate the detuning over time, with a precise calibration of the zero-detuning (\textit{i.e.} position of the secondary color comb tooth).
We note that we obtain the absolute value of the detuning and need to process to change its sign based on the direction of the triangular waveform applied to the CTL piezo element. From knowledge of the CTL characteristics, we know that a high-to-low voltage results in a high-to-low frequency, hence the detuning is negative for times before the calibration frequency reaches zero, since the oscilloscope is triggered for positive slope with the piezo ramp signal.

\section{Laser phase noise measurement
\label{supsec:phase_noise}}

\begin{figure*}[h]
    \centering
    \includegraphics{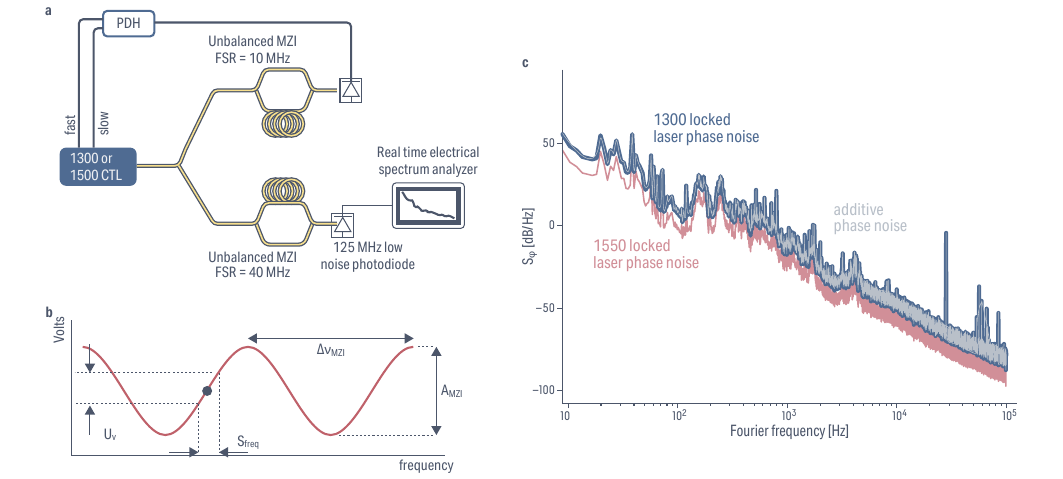}
    \caption{\label{figsup:freq_noise}
    \textbf{Phase noise characterization of the auxiliary and reference pump}. %
    \textbf{a} Schematic of the experimental setup to measure the auxiliary or reference laser phase noise. We lock the laser to an unbalanced fiber Mach-Zehnder interferometer (MZI) with a 10~MHz free spectral range. We measure the phase noise of the laser using another 40~MHz MZI, and a low noise 125~MHz photodiode. %
    \textbf{b} Schematic explaining the relationship between the noise measured with the 40~MHz FSR MZI and the laser frequency noise.
    \textbf{c} Phase noise of the 1300~nm laser (blue) and 1550~nm laser (red). The phase noise of the 1300~nm laser is larger than the 1550~nm laser, and if accounting for additive noise between them (gray) shows that it is predominantly noise from the 1300~nm laser. The power is referenced to that of the carrier, namely, dBc/Hz%
    }
\end{figure*}

We lock the 1300~nm and 1500~nm CTLs using independent 10~MHz free spectral range (FSR) Mach-Zenhder interferometers (MZIs), and using a Pound-Drever-Hall technique allows us to actuate the current and piezo element of the laser (respectively fast and slow feebdack) to lock to the side of a fringe of the MZI [\cref{figsup:freq_noise}]. To measure their respective phase noise, we use another MZI with a 40~MHz FSR. Setting the laser to be at a quadrature point of this MZI, we measure the residual MZI noise using a real time electrical spectum analyzer (RSA). Understanding the noise obtained from the MZI [\cref{figsup:freq_noise}b], we can retrieve the frequency noise of the laser following:

\begin{align}
    S_\mr{\nu} = \left[\frac{\Delta\nu_\mr{MZI}}{\pi} \sin^{-1}\left(\frac{U_\mr{MZI}}{A_\mr{mzi}} \right)\right]^2 \quad \mr{[Hz^2/Hz]} \nonumber
\end{align}

\noindent with $U_\mr{MZI} = \sqrt{50 \times 10^{-3} \times 10^{S_\mr{MZI}/10}}$ the voltage obtained from spectral measurement, where $S_\mr{MZI}$ is the optical power in decibels normalized to 1~mW of the MZI noise at quadrature (i.e., in dBm), the spectrum analyzer impedance is assumed to be 50~$\Omega$, $A_\mr{mzi} = 584$~mV is the amplitude of the sinusoidal modulation with frequency of the MZI, and $\Delta \nu_\mr{MZI}=40$~MHz is the free spectral range of the MZI used for noise measurement. 

We can obtain the phase noise from the frequency noise, where the power is referenced to that of the carrier, namely, dBc/Hz: 
\begin{align}
    S_\varphi = 10\log_{10}\left(\frac{S_\mr{\nu}}{\nu^2}\right) \quad \mr{[dB/Hz]}\nonumber
\end{align}

When comparing the phase noise of the two locked lasers, the 1500~nm laser exhibits lower noise~\cref{figsup:freq_noise}. When accounting for additive noise since the two lasers are locked to independent MZIs, as the 1300~nm laser being more noisy than the 1500~nm laser, the noise of the 1300~nm laser is predominant. We therefore consider OFD of the 1300~nm laser noise when assessing the repetition rate noise under color-KIS.

\section{Leveraging the cooler as an auxiliary pump for counter-propagating color-KIS}
\begin{figure*}[h]
    \centering
    \includegraphics{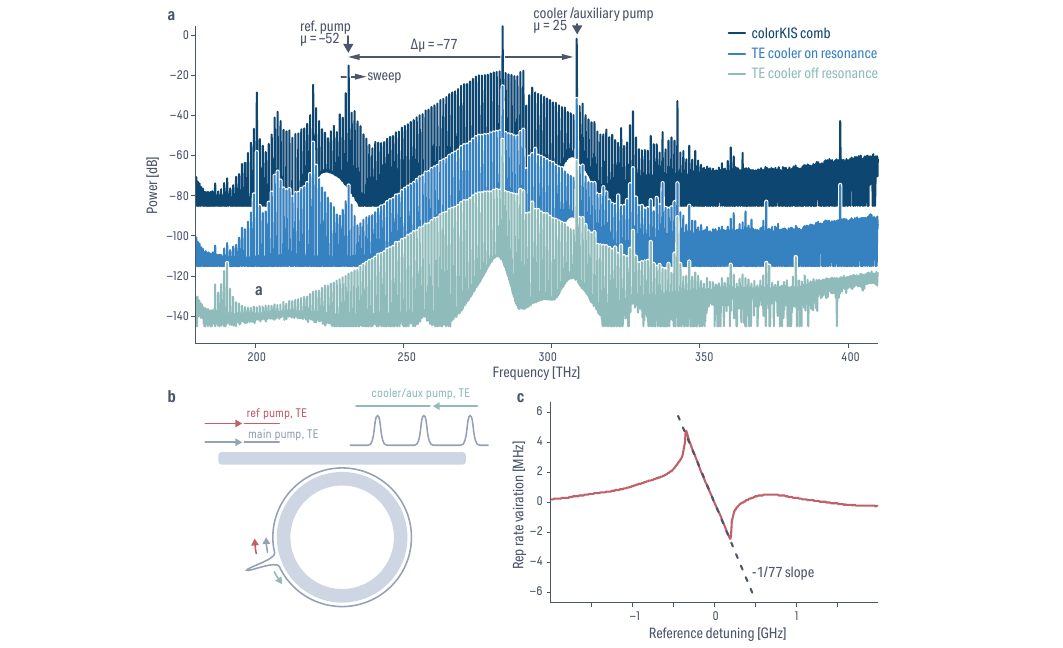}
    \caption{\label{figsup:KIScooler}
    \textbf{Counter-propagating cooler pump used as the auxiliary pump for synthetic DW creation and counter-propagating color-KIS}. %
    \textbf{a} Optical frequency comb with the transverse electric counter-propagating cooler set off resonance such that its induced nonlinearity is minimized (foreground spectrum in teal). When the cooler is set to be on resonance, it acts as an auxiliary pump that creates a new set of colors with the synthetic DWs at $\mu = -52$ and $\mu = -82$, while through cascading a third color is also created with other synthetic DWs (middle spectrum in light blue). Hence, we can use the reference laser at the synthetic DW created by the cooler/auxiliary pump laser, enabling color-KIS with a limited number of lasers (background spectrum in dark blue). %
    \textbf{b} Schematic of the counter-propagating color-KIS scheme. Here, the cooler laser also acts as an auxiliary pump that mediates the creation of a new soliton color and synthetic dispersive waves. Instead of using the cooler in the same propagation direction as the reference and main pumps, we can leverage that since KIS is XPM-mediated, it can even be triggered in a counter-propagating fashion.
    \textbf{c} The repetition rate entrainment follows the slope defined by the cooler and auxiliary pump mode spacing, which, since the cooler is of such high power, enables a larger OFD than presented in the main text.  Here, the auxiliary/cooler pump and reference pump are on opposite sides of the main pump, whereas in the main text they were set with a $\mu$ of the same sign.
    }
\end{figure*}

In order to demonstrate that color-KIS can occur with a limited set of lasers, we leverage the fact that we already use a cooler pump to thermally stabilize the microring resonator for adiabatic access to the soliton. While in the main text of the manuscript we use a cross-polarized cooler pump (transverse magnetic polarization) relative to the main pump (transverse electric polarization) to minimize the nonlinearity induced by the cooler, here we instead operate in the same polarization to allow the cooler to also act as an auxiliary pump that creates a second soliton color and new set of DWs. Since the cooler is high power, even synthetic DWs on the opposite side of the pump than the reference (i.e., a different sign for $\mu$) and which are usually inefficient can be created~[\cref{figsup:KIScooler}a], enabling a larger OFD factor than presented in the main text. Interestingly, this effect still occurs despite the cooler remaining in a counter-propagating direction relative to the main and reference pump [\cref{figsup:KIScooler}b]. As KIS is mediated through cross-phase modulation, counter-propagating KIS can happen, as previously demonstrated  between two counter-propagating solitons~\cite{YangNaturePhoton2017}, where the synchronization is mediated by coherent backscattering, which we believe is also the case in our work. However, in linear transmission measurements we do not observe lifting of the resonance degeneracy between the two traveling wave modes at either the reference mode or at the cooler mode, and hence the synchronization could be mediated by a relatively weak backscattering relative to the resonance linewidth, or by Rayleigh scattering at other modes where the secondary color exists. Regadless, we leverage the resulting synthetic dispersive wave for large OFD, while also simplifying the architectural complexity of color-KIS by reducing the number of required lasers. We send a reference at $\mu=-52$ while the cooler is present at $\mu = + 25$, enabling an OFD of $\Delta\mu = -77$, which is confirmed by repetition rate entrainment measured at the DKS comb component [\cref{figsup:KIScooler}c]. 

\end{document}